\documentclass[11pt,twoside]{jgsp}

\def\aut#1#2{
{\small
\noindent
\parbox[t]{6.5cm}{#1}
 \hfill
\parbox[t]{6.5cm}{#2}
}
}

\begin{document}

\thispagestyle{plain}

\title{Spin Networks in Quantum Gravity}

\author{Miguel Lorente}

\date{}

\maketitle

\begin{abstract}
This is a review paper about one of the approaches to unify Quantum Mechanics and the theory of General Relativity. Starting from the pioneer work of Regge and
Penrose other scientists have constructed state sum models, as Feymann path integrals, that are topological invariant on the triangulated Riemannian surfaces, and
that in the continuous limit become the Hilbert-Einstein action.
\end{abstract}

\label{first}

\section[]{Introduction}

This is a review talk to present the main ideas of some line of research in quantum gravity, namely the spin foam approach, that has been explored by a
great number of physicists and mathematicians and has attracted much attention. The three main lines of research in quantum gravity are denoted as
``canonical'', ``covariant'' or ``sum over histories'' \cite{rov}

The canonical line of research is a theory in which the Hilbert space carries a representation of the quantum operators corresponding to the full metric
without background metric to be fixed. It can be considered as a quantum field theory on a differentiable manifold. The basis of the Hilbert space are
cilindrical functions defined on a graph (Wilson loops) depending on Ashtehar variables \cite{r-u}. A very important result of this approach was the discrete
eigenvalues for the area and volumen operators.

The covariant line of research is the attempt to built the theory as a quantum field theory of the fluctuations of the metric over a flat Minkowski space,
or some other background metric space. The theory has been proved to be renormalizable and finite order by order. \cite{a-l}

The sum over histories line of research uses the Feymann path integral to quantize the Einstein Hilbert action. There exist a duality between this model and group
field theories. The sum over spin foam can be generated as the Feymann perturbative expansion of the group field theories. Each space-time appears as the Feymann
graph of the auxiliary groups field theory. \cite{bae} Our presentation is going to be concentrated on this third line of research, namely, the spin network and the spin
foam models, from an historical point of view.

\section{Regge calculus (1961) }
The  Regge's paper \cite{reg} was a pionnier work in the discretization of GR, that was motivated by the need to avoid coordinates, because the physical prediction of the
theory was coordinates independent. For that purpose he discretizes a continuous manifold by $n$-simplices, that are glued together by identification of their
$(n-1)$-simplices. The curvature lies on the $(n-2)$- dimensional subspaces, known as hinges or bones. From pedagogical reasons we take a triangulation of a
2-dimensional surface. When a collection of triangle meeting at a vertex is flattened there will be a gap or deficit angle $\epsilon$, indicating the presence of
curvature. Using the Gauss-Bonet formula we can calculate the excess angle by $\epsilon=KA$, where $K$ is the curvature at that vertex and $A$ the area of the
triangles around the vertex. If the number of vertices increases we can take $K=\epsilon \rho$ , where $\rho$ is the density of vertices in the triangulation (=
number of vertices by unit area). This method is easily enlarged to higher dimensions. 

In order to have connecton with GR we traslate into the triangulated surface (the skeleton) the Hilbert-Einstein action ${\cal L} = (1/8\pi )\int {Kd^4 x\sqrt { - g}
}
$ where $K$ is the scalar curvature in 4-dimensions. The discrete version for a 4-dimensional skeleton is given in terms of the deficit angle in each bone where the
curvature $K$ is calculated and some measure function $L$ is defined:
$${\cal L} = \sum\limits_{n = 1}^N {\epsilon _n L_n } $$
here the summation extends to all the bones in the skeleton. In the continuous case the Einstein's equations are derived from a stationary action, varying ${\cal L}$
with respect to the metric. In the discrete version one derives the action with respect to the edge lengths, because in the simplicial decomposition all the
properties can be derived from these edges. Using Schlaefli differential identity one finds
$$
\delta {\cal L} = {1 \over {8\pi }}\sum\limits_{n = 1}^N {\varepsilon _n \delta L_n }  = 0\quad  \Rightarrow \quad \sum\limits_{n = 1}^N {\varepsilon _n {{\partial
L_n } \over {\partial l_p }} = o} 
$$
which is the discrete version of Einstein's equations \cite{r-w}

\section{The Ponzano-Regge model (1968) }
Some years later Ponzano and Regge \cite{p-r} made use of $\{6j \}$ symbols attached to the tetrahedra decomposition in order to calculate the state sum and were able to
connect it to the Feymann integral corresponding to the Hilbert-Einstein action.

The $\{6j \}$ Wigner simbols is a generalization of the Clebsch-Gordan coefficients that appear in the coupling of two angular momenta $J = J_1  + J_2. $

The new basis is given in term of the old basis:
$$
\left| {j_1 j_2 jm} \right\rangle  = \sum {\left\langle {j_1 j_2 m_1 m_2 } \right|\left. {j_1 j_2 jm} \right\rangle } \left| {j_1 j_2 m_1 m_2 } \right\rangle 
$$
If we couple $J$ with a new angular momentum $J_3$ we have two possibilities
$$
\left( {J_1  + J_2 } \right) + J_3  = J\quad {\rm or} \quad J_1  + \left( {J_2  + J_3 } \right) = J
$$
In the first case the new basis is given (in obvious notation)
$$
\left| {j_1 j_2 j_3 j_{12} jm} \right\rangle  = \sum {\left\langle {j_1 j_2 j_3 m_1 m_2 m_3 } \right|\left. {j_1 j_2 j_3 j_{12} jm} \right\rangle } \left| {j_1 j_2 j_3 m_1 m_2 m_3 } \right\rangle 
$$
In the second case the new basis is given by
$$
\left| {j_1 j_2 j_3 j_{23} jm} \right\rangle  = \sum {\left\langle {j_1 j_2 j_3 m_1 m_2 m_3 } \right|\left. {j_1 j_2 j_3 j_{23} jm} \right\rangle } \left| {j_1 j_2 j_3 m_1 m_2 m_3 } \right\rangle 
$$
The transformating matrix between the two basis is given precisely by the $\{6j \}$ symbols, namely,
$$
U\left( {j_{12} j_{23} } \right) = \left( { - 1} \right)^{j_1  + j_2  + j_3  + j} \sqrt {\left( {2j_{12}  + 1} \right)\left( {2j_{23}  + 1} \right)} \left\{
 \begin{matrix}
   \begin{array}{ccc}{j_1 } & {j_2 } & {j_{12} }  \cr 
   {j_3 } & j & {j_{23} }  \cr 
\end{array} 
\end{matrix} 
 \right\}
$$

Given a tetrahedra decomposition of a 3-dimensional surface we can attach a $\{6j \}$ symbol to each tetrahedra, the edges of which have the length equal to the
numerical values of the angular momenta appearing in the $6j$-symbol.

\begin{center}
\includegraphics{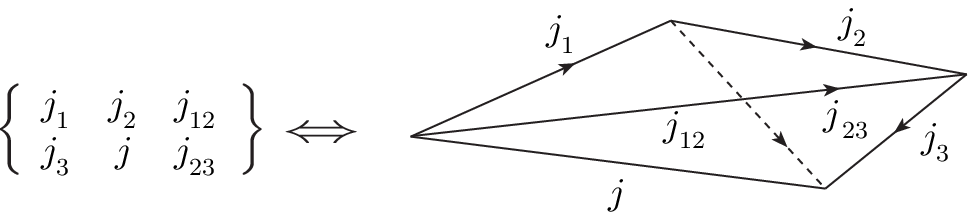}
\end{center}

This choice is consistent with the inequalities
$$
j_{12}  < j_1  + j_2 \quad {\rm and} \quad j_{23}  < j_2  + j_3 
$$
and the equalities 
$$
j_1  + j_2  + j_{12}  \subset N,\quad j_2  + j_3  + j_{23}  \subset N
$$
The $\{6j \}$ symbols are proportional to the Racah polynomials \cite{n-s-u}
$$
\left( { - 1} \right)^{j_1  + j_2  + j_3 } \sqrt {\left( {2j_{12}  + 1} \right)\left( {2j_{23}  + 1} \right)} \left\{ 
\begin{matrix}
   \begin{array}{ccc}{j_1 } & {j_2 } & {j_{12} }  \cr 
   {j_3 } & j & {j_{23} }  \cr 
\end{array}
\end{matrix}
 \right\} = {{\sqrt {\rho (x)} } \over {d_n }}U_n^{(\alpha ,\beta )} (x,a,b)
$$

From this equality and the assymptotic properties of Racah polynomials one can derive very important limit
$$
 \left\{ 
\begin{matrix}
 \begin{array}{ccc} {j_1 } & {j_2 } & {j_{3} }  \cr 
   {j_4 } & {j_5 } & {j_6 }  \cr 
\end{array}
\end{matrix}
 \right\}\mathrel{\mathop{\kern0pt\longrightarrow}
\limits_{j_i  \to \infty }} {1 \over {\sqrt {12\pi V} }}\cos \left\{ {\sum\limits_{i = 1}^6 {\left( {j_i + {1 \over 2}} \right)} \vartheta _i  + {\pi  \over 4}}
\right\}
$$
where $V$ is the volume of the tetrahedra and $\vartheta _i$ the exterior dihedral angle adjoint to the edge $j_i$
In order to see the conection between $\{6j \}$ symbols and quantum gravity we take a tetrahedra decomposition and take external edges $l_i$ of the bounding surface
and internal edges $x_i$. Then Ponzano and Regge construct the state sum as follows

$$
S\left( {l_i } \right) = \sum\limits_{x_i } {\prod\limits_{\rm tetrahedra} {\left\{ {6j} \right\}\left( { - 1} \right)^X \prod\limits_{\rm edges} {\left( {2x_i  + 1}
\right)} } } 
$$
where $X$ is a phase factor. Substituting the $\{6j \}$ symbols by their assymptotic values and the function cosine by the Euler expression we arrive at

$$
S\left( {l_i } \right) = \sum\limits_{x_i } {\prod\limits_{i = 1} {\left( {2x_i  + 1} \right)\exp \left\{ {i\left[ {\left( {\sum\limits_{{\rm tetrahedra} \,k}
{\vartheta _i^k } } \right) - \pi p_i  + 2\pi } \right]x_i } \right.} } 
$$

We may replace the summation with an integral. Then the most important contribution comes from the stationary phase that is to say when $\sum\limits_{\rm tetrahedra}\left( {\pi  - \vartheta _i^k } \right) = 2\pi $

Introducing this value in the state sum we obtain \cite{r-w}
$$
S\left( {l_i } \right) = \int {\prod\limits_{x_i } {\left( {2x_i  + 1} \right)dx_i \exp \left( {i\sum {j_l \epsilon _l } } \right)} } 
$$
where $\epsilon _{\ell}  = 2\pi  - \sum\limits_k {\left( {\pi  - \theta _l^k } \right)}$

The summation $\sum {j_l \epsilon _l } $ approaches the Hilbert-Einstein action that was given in the Regge calculus, therefore in the limit the state sum
strongly resembles a Feymann summation over histories with density of Lagrangian 
${\cal L} = \left( {1/8\pi } \right)  \int Ra^4 x $ $\sqrt { - g} $\quad namely
$$
S = \int {d\mu (x_i } )e^{i{\cal L}} 
$$

\section{Penrose's spin networks (1971) }
Penrose was interested in the interpretation of space-time \cite{pen} by purely combinatorial properties of some elementary units that are connected among themselves by
some interactions that follow the angular momentum quantum rules, and form some network of elementary units with assigned spins. Soon it was realized that the spin
network was analog to simplicial quantum gravity, in particular the Ponzano-Regge model \cite{h-p}. His networks had trivalent vertices and the edges were labeled with
spin, satisfying the standard conditions at the vertices. The model was generalized to any group different from the rotation group. Formaly a spin network ia a
triple $\left( {\gamma ,\rho ,\iota } \right)$ where 
\begin{enumerate}
\item [i)]  
$\gamma$ is a graph with a finite set of edges $e$ and a finite set of vertices, $v$,

\item [ii)]  to each edge $e$ we attach an irrep of a group $G$, $\rho_e$

\item [iii)]  to each vertex $v$  we attach an intertwiner.
\end{enumerate}
When we take the dual of an spin network we obtain a triangulated figure, which, after embedding in a 3-dimensional manifold becomes the triangulation of Regge
calculus.

\section{The Turaev-Viro state-sum invariant (1992) }
They defined a state sum for triangulated 3-manifold (as in the Ponzano-Regge model) that was independent os the triangulation and finite \cite{t-v}. For this purpose they
assign a value from the set $I_r  = \left( {0,1/2,1,\,\left( {r - 2} \right)/2} \right)$, integer, to each edge of the triangulation, subject to the condition that
the ``coloring'' of the three edges forming a triangle should satisfy the triangle inequalities and their sum should be and integer less than or equal to $r-2$.
Define the quantum object
$$
\left| M \right|_\phi   = \omega ^{ - 2\rho } \prod\limits_{{\rm tetra}\,k} {\left| {T_k^\varphi  } \right|} \prod\limits_{{\rm edge}\,j} {\omega _j^2 } 
$$
where $\varphi$ is an admisible coloring of the edge $j$,
$$
\omega _j  = \left( { - i} \right)^j \left[ {2j + 1} \right]^{1/2} \quad ,\quad \omega _{}^2  = \sum\limits_{j \in I_r } {\omega _j^4 } $$ and $\left| {T_k^\varphi 
} \right|$ is the quantum $6j$-simbol corresponding to the tetrahedron $k$ with coloring $\varphi $, such that
$$
\left| 
\begin{matrix}
  \begin{array}{ccc} i & j & k  \cr 
   l & m & n  \cr 
\end{array}
\end{matrix}
  \right| = \left( { - 1} \right)^{i + j + k + l + m + n} \left\{ 
\begin{matrix}
  \begin{array}{ccc}  {\left[ i \right]} & {\left[ j \right]} & {\left[ k \right]}  \cr 
   {\left[ l \right]} & {\left[ m \right]} & {\left[ n \right]}  \cr 
\end{array}
\end{matrix}
 \right\}
$$
where [$n$] is the quantum number satisfying $\left[ n \right] \to n$. Summing $\left| M \right|_\phi$ over all admisible coloring we obtain an expresion in the
limit $q \to 1$ or $r \to \infty $ that becomes identical to the Ponzano-Regge state sum. Turaev and Viro proved that their expression is manifold invariant (or
independent of triangulation) under Alexander moves, and also finite.

\section{The 3-dimensional Boulatov model (1992) }
The Ponzano-Regge state sum and the Turaev-Viro model are defined over 3-dimensional manifold. To enlarge the model to four dimensions it vas necessary to increase
the Wigner symbols to $3nj$. The key to this approach was given by Boulatov \cite{bou} by the use of topological lattice gauge theories, taking group elements as variables
(matrix models). The basic objects is a set of real functions of 3 variables $\phi \left( {x,y,z} \right)$ (where $x,y,z \in SU(2))$ invariant under simultaneous
right shift of all variables by $u \in SU(2)$ and also by cyclic permutation of $x,y,z$. This function $\phi$ can be expanded, by Peter-Weyl theorem, in terms of
representations of $SU(2)$ and 3j-symbols. An action of interest can be constructed with those functions as follows
\begin{eqnarray*}
S &=& {1 \over 2}\int {dxdydz\phi ^2 \left( {x,y,z} \right)} - \\
&-&{{\lambda \over {4!}}} \int {dxdydzdudvdw} \phi \left( {x,y,z} \right)\phi \left( {x,u,v} \right)\phi \left(
{y,v,w} \right)\phi \left( {z,w,u} \right)
\end{eqnarray*}
If we attach the variable to the edges, the first term (the kinematical term) represent two glued triangle and the second one (the interacting term) four triangles
forming a tetrahedron. Substituting the Fourier expansion of funcion $\phi $, and integrating out group variables we obtain an action depending on the Fourier
coefficientes and 6j-simbols. From this result we calculate the partition function as a Feymann path integral with respect to the Fourier coefficients
$$
Z = \int {D\phi e^{ - S}  = \sum\limits_{\left\{ C \right\}} {\lambda ^{N_3 } } \sum\limits_j {\prod\limits_l {\left( {2j_e  + 1} \right)\prod\limits_T {\left\{ {6j} \right\}} } } } 
$$
where the products extend to all tetrahedra T, all edges $l$, and the summation extend to all the representations $\left\{ j \right\}$, all the simplicial complexes
$\left\{ C \right\}$ and $N_3$ is the number of tetrahedra in complex $C$. This partition function is equivalent, up to renormalization, to Ponzano-Regge state sum
applied to some triangulation of 3-dimensional manifold. The underlying mathematical structure is a topological lattice gauge theory, it has the advantage that is
topological invariant. In order to prove it, Boulatov used the Alexander moves, by which one complex, and the corresponding partition function is topological
invariant

\section{The 4-dimensional Ooguri's model (1992) }
The 3-dimensional Boulatov model paved the way to the Ooguri's model in four dimensions. \cite{oog} Let $\phi $ be a real valued function of four variables 
$\phi \left( {g_1 ,g_2 ,g_3 ,g_4 } \right)$ on $G\left( {g_i  \in G} \right)$ a compact group. For simplicity we take $G=SU(2)$. We require $\phi $ to
be invariant under the right action of $G$, and by cyclic permutation of these variables. The Peter-Weyl theorem, we can expand $\phi $  in terms of these
representations and the 3j-symbols. We define the action 
\begin{eqnarray*}
S &=& {1 \over 2}\int {\prod {dg_i \phi ^2 \left( {g_1 g_2 g_3 g_4 } \right)} }  + {\lambda  \over {5!}}\int {\prod\limits_{i = 1}^{10} {dg_i \phi } } \left( {g_1 g_2
g_3 g_4 } \right) \times
\\ &\times& \phi \left( {g_4 g_5 g_6 g_7 } \right)\phi \left( {g_7 g_3 g_8 g_9 } \right)\phi \left( {g_9 g_6 g_2 g_{10} } \right)\phi \left( {g_{10} g_8 g_5 g_1 }
\right)
\end{eqnarray*}
where the first term (the kinematical term) represents the coupling of a tetrahedrum with itself because each element $g_i$ is associated to each face of the
tetrahedrum, and the second term (the interacting term) represents gluing faces of five tetrahedra to make a four-simplex. Substituting the Fourier expansion into
the action we can integrate out the group variable, and then the action can be used to calculate a partition function as a Feyman path integral with respect to this
action:
$$
Z = \int {DMe^{ - S\left( M \right)} }  = \sum\limits_C {\lambda ^{N_4 } } \sum\limits_{\left\{ j \right\}} {\prod\limits_t {\left( {2j_t  + 1} \right)\prod\limits_T {\left\{ {6j} \right\}\prod\limits_S {\left\{ {15j} \right\}} } } } 
$$
where the integral is defined in terms of the Fourier coefficients $M$, appearing in the action and in the measure, the first sum is over all complexes $C$
(four-dimensional combinatorial manifolds), $N_4(C)$ is the number of 4-simplices in $C$, the second summation is over all irreducible representations os $SU(2)$
with angular momentum $j; t, T$ and $S$ are the triangles, tetrahedra and 4-simplexes respectively apearing in the complex. Ooguri also proved that the partition
function is topological invariant under the Alexander moves. As in the Boulatov model two complexes are combinatorially equivalent if and only if they are connected
by a sequence of transformations called Alexander moves.

\section{The Barrett-Crane model (1998) }
A more abstract approach was taken by Barrett and Crane \cite{b-c1} generalizing Penrose's spin networks to four dimensions. The novelty of this model consists in the
association of representations of $SO(4)$ group to the faces of the tetrahedra, instead of the edges. They decompose a triangulation of a
4-dimensional manifold into 4-simplices, the geometrical properties of which are characterized in terms of bivectors.

A geometric 4-simplex in Euclidean space is given by the embedding of an ordered set of 5 points in $R^4(0,x,y,z,t)$ which is
required to be non-degenerate (the points should not lie in any hyperplane). Each triangle in it determines a bivector
constructed out of the vectors for the edges. Barrett and Crane proved that classically, a geometric 4-simplex in Euclidean space
is completely characterized (up to parallel translation an inversion through the origin) by a set of 10 bivectors $b_i$, each
corresponding to a triangle in the 4-simplex and satisfying the following properties:
 \begin{enumerate}
\item [i)]  
the bivector changes sign if the orientation of the triangle is changed;

\item [ii)] each bivector is simple, i.e. is given by the wedge product of two vectors for the edges;

\item [iii)]  if two triangles share a common edge, the sum of the two bivector is simple;

\item [iv)]  the sum (considering orientation) of the 4 bivectors corresponding to the faces of a tetrahedron is zero;

\item [v)]  for six triangles sharing the same vertex, the six corresponding bivectors are linearly independent;

\item [vi)]  the bivectors (thought of as operators) corresponding to triangles meeting at a vertex of a tetrahedron satisfy tr $b_1\left[
{b_2,b_3}
\right]>0$ i.e. the tetrahedron has non-zero volume.
 \end{enumerate}
Then Barrett and Crane define the quantum 4-simplex with the help of bivectors thought as elements of the Lie algebra $SO(4)$,
associating a representation to each triangle and a tensor to each tetrahedron. The representations chosen should satisfy the
following conditions corresponding to the geometrical ones:
\begin{enumerate}
\item [i)] different orientations of a triangle correspond to dual representations;

\item [ii)] the representations of the triangles are ``simple'' representations of $SO(4)$, i.e. $j_1=j_2$;

\item [iii)] given two triangles, if we decompose the pair of representations into its Clebsch-Gordan series,
the tensor for the tetrahedron is decomposed into summands which are non-zero only for simple representations;

\item [iv)] the tensor for the tetrahedron is invariant under $SO(4)$. 

Now it is easy to construct an amplitude for the quantum
4-simplex. The graph for a relativistic spin network is the 1-complex, dual to the boundary of the 4-simplex, having five
4-valent vertices (corresponding to the five tetrahedra), with each of the ten edges connecting two different vertices
(corresponding to the ten triangles of the 4-simplex each shared by two tetrahedra). Now we associate to each triangle (the
dual of which is an edge) a simple representation of the algebra $SO(4)$ and to each tetrahedra (the dual of which is a
vertex) a intertwiner; and to a 4-simplex the product of the five intertwiner with the indices suitable contracted, and the
sum for all possible representations. The proposed state sum suitable for quantum gravity for a given triangulation
(decomposed into 4-simplices) is 
 \vspace{-0,2cm}
\end{enumerate}
\begin{eqnarray*}
Z_{BC}=\sum\limits_J {\prod\limits_{{\rm triang.}} {A_{{\rm tr}}}}\prod\limits_{{\rm tetrahedra}} {A_{{\rm
tetr.}}}\prod\limits_{4-{\rm simplices}} {A_{{\rm simp.}}}
\end{eqnarray*}
where the sum extends to all possible values of the representations $J$.

\begin{center} 
\includegraphics{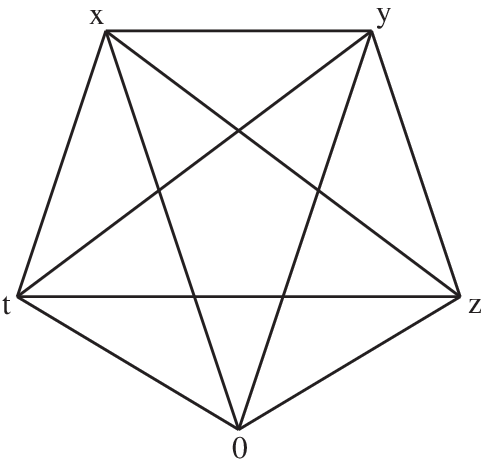}
\end{center}

In order to know the representation attached to each triangle of the tesselation, we take the unitary representation of SO(4) in terms of Euler angles.
$$
U\left( {\varphi ,\theta ,\tau ,\alpha ,\beta ,\gamma } \right) = R_3 \left( \varphi  \right)R_2 \left( \theta  \right)S_3 \left( \tau  \right)R_3 \left( \alpha  \right)R_2 \left( \beta  \right)R_3 \left( \gamma  \right)
$$
where $R_2$ is the rotation matrix in the $(x_1x_3)$ plane, $R_3$ the rotation matrix in the $(x_1x_2)$ plane and $S_3$ the rotation (``boost'') in the $(x_3x_4)$ 
plane. In the angular momentum basis, the action of $S_3$ is a follows
$$
S_3 \left( \tau  \right)\psi _{jm}  = \sum\limits_{j'} {d_{j'jm}^{j_1 j_2 } \left( \tau  \right)\psi _{j'm} } 
$$
where
$$
d_{j'jm}^{j_1 j_2 } \left( \tau  \right) = \sum\limits_{m_1  + m_2  = m} {\left\langle {{j_1 j_2 m_1 m_2 }}
 \mathrel{\left | {\vphantom {{j_1 j_2 m_1 m_2 } {jm}}}
 \right. \kern-\nulldelimiterspace}
 {{jm}} \right\rangle } e^{ - i\left( {m_1  - m_2 } \right)\tau } \left\langle {{j_1 j_2 m_1 m_2 }}
 \mathrel{\left | {\vphantom {{j_1 j_2 m_1 m_2 } {j'm}}}
 \right. \kern-\nulldelimiterspace}
 {{j'm}} \right\rangle 
$$
is the Biedenharn-Dolginov function \cite{bie}

\section{Evaluation of the state sum for the 4-dimensional spin network }
In order to evaluate the state sum for a particular triangulation of the total $R^4$ space by 4-simplices, we assign an element $h_k  \in SU(2)$ to each tetrahedrum
of the 4-simplex $(k=1,2,3,4,5)$ and a representation $\rho _{kl} $ of $SO(4)$ to each triangle shared by two tetrahedra. From this triangulation we obtain an
2-complex by the dual graph where one vertex corresponds to a tetrahedrum and an edge corresponds to triangle, with the ends of the edges identified with the
vertices. Then we attach a representation of $SU(2)$, $\rho \left( {h_k } \right)$ and $\rho \left( {h_l } \right)$ to the vertices $k$ and $l$ and contract both
representation along the edges $(k,l)\equiv e$, giving
$$
Tr\rho \left( {h_k } \right)\rho \left( {h_l^{ - 1} } \right) = Tr\rho _{kl} \left( {h_k h_l^{ - 1} } \right)
$$
where $\rho _{kl}$ is the representation of $SO(4)$ corresponding to the product $h_k h_l^{ - 1}$, the left and right components of the $SO(4)$ group. The state sum
for the 2-dimensional complex (the Feymann graph of the model) is obtained by taking the product for all the edges of the graph and integrating for all the copies of
$SU(2)$
$$
I = \int\limits_{h \in SU(2)^5 }^{} {\prod\limits_{} {Tr\rho _{kl} \left( {h_k h_l^{ - 1} } \right)} } 
$$
Due to the trace condition this expression is invariant under left and right multiplication of some elements of $SU(2)$. \cite{bar}

For the representation $\rho_{kl}$ we choose the spherical function with respect to the identity representation. Given a completely irreducible
representation of the group $G:g \to T_g$ on the space $R$, we define the spherical function with respect to the finite irrepr. of the subgroup $K$
$$
f_k (g) = Tr\left\{ {E^k T_g } \right\}
$$
where $E^k$ is a projector of $R$ onto the space $R_k$ of $K$ 

We take for $G\equiv SO(4)$ the simple representation $(j_1=j_2)$ and for the subgroup $SU(2)$ the identity representation $k=0$.  Since $f_k$ is invariant under $K$
we can restrict the unitary representations to those of the boost $S_3(\tau)$. With the help of the Biedenharn-Dolginov function it can be proved
$$
f_0 \left( \tau  \right) = Tr\left\{ {E^0 S_3 \left( \tau  \right)} \right\} = {{\sin \left( {2j_1  + 1} \right)\tau } \over {\sin \tau }}
$$
With this formula it is still possible to give a geometrical interpretation of the probability amplitude encompassed in the trace. In fact the spin dependent factor
appearing in the exponential of the spherical function
$$
e^{i\left( {2j_{kl}  + 1} \right)\tau _{kl} } 
$$
corresponding to the two tetrahedra $k,l$ intersecting the triangle $kl$, can be interpreted as the product of the angle between the two vectors $h_k,h_l$,
perpendicular to the triangle, and the area $A_{kl}  = 2j_{kl}  + 1$ of the intersecting triangle, $j_{kl}$ being the spin corresponding to the representation
$\rho_{kl}$ associated to the triangle $k,l$.
Substituting this value in the state sum, we obtain 
$$
I = \prod\limits_{h \in SU(2)} {{1 \over {\sin \tau _{kl} }}\exp \left( {i\sum\limits_{{\rm triangle}kl} {A_{kl} \tau _{kl} } } \right)} 
$$ 
where the product extend to all tetrahedra with the vector $h$ perpendicular to the subspace where the tetrahedra is embedded, and summation is extended to all the
triangle $k,l$ intersected by two tetrahedra $k$ and $l$. The exponential term correspond to the Regge action, that in the assymptotic limit becomes the
Hilbert-Einstein action \cite{b-w}

Because we are interested in the physical and mathematical properties of the Barrett-Crane model, we mention some recent work about this model combined with the
matrix model approach of Boulatov and Ooguri.\cite{d-r-k} In this work the 2-dimensional quantum space-time emerges as a Feymann graph, in the manner of the 4 dimensional
matrix models. In this way a spin foam model is connected to the Feyman diagram of quantum gravity.

\section{The Lorentzian spin foam model }
Now we apply the same technique to calculate the state sum invariant under the Lorentz group that we have used in the case of the $SO(4)$ group for the Barrett-Crane
model

The unitary irreducible representation of the $SL(2,C)$ group for the principal series is given by the formula [7]
$$
\left( 
{T_g^{\left[ {m,\rho } \right]} \psi } \right)\left( z \right) = \left( {\beta z + \delta } \right)^{m + i\rho  - 2} \left( {\beta z + \delta
} \right)^{ - m} \psi \left( {{{\alpha z + \gamma } \over {\beta z + \delta }}} \right)
$$
where $g = \left( 
\begin{matrix}
   \begin{array}{cc}\alpha  & \beta   \cr 
   \gamma  & \delta   \cr 
\end{array}
\end{matrix}
  \right) \in SL(2,C)$, $m$, integer, $\rho$ ral $\psi \left( z \right) \in L^2 \left( \mathbb C \right)$. The numbers $m, \rho$ determine the eigenvalues of the
representation
$$
C_1  =  - {{m^2  - \rho ^2  - 4} \over 2}\quad ,\quad C_2  = m\rho 
$$
In order to calculate the state sum we need the spherical functions of the irrep of $SL(2,C)$. These are given in terms of the Biedenharn-Dolginov function that
correponds to the boost operator
\begin{eqnarray*}
d_{JJ'M}^{\left[ {m,\rho } \right]} \left( \tau  \right) &= \int\limits_{ - \infty }^\infty  {d_{J - M}^{ - 1} p_{J - M}^{\left( {M - m,M + m)} \right)} } \left(
{\lambda ,\rho } \right)e^{ - i\tau \lambda }  \times \\
 & \times\;d_{J' - M}^{ - 1} p_{J' - M}^{\left( {M - m,M + m)} \right)} \left( {\lambda ,\rho } \right)\omega \left( \lambda  \right)d\lambda 
\end{eqnarray*} 
where $J,M$ are the angular momentum eigenvalues, $d_n$ is some normalization constant, and $p_n^{\left( {\alpha ,\beta } \right)}$ are the Hahn polynomials of
imaginary argument \cite{n-s-u}. Given the unitary representation $T_g$ of the group $SL(2,C)$ and the identity representation of $SU(2)$, the spherical functions is defined
as in the case of $SO(4)$
$$
f_0 \left( \tau  \right) = Tr\left\{ {E^0 T_g } \right\} = d_{000}^{0,\rho } \left( \tau  \right) = {1 \over \rho }{{\sin \rho \tau } \over {sh\tau }}
$$
where the last step has been calculated with the residue theorem.

\section{A SO(3,1) invariant for the state sum of spin foam model} 

As in the case of euclidean $SO(4)$ invariant model, we take a non degenerate finite triangulation of a 4-manifold. We consider
the 4-simplices in the homogeneous space $SO(3,1)/SO(3)$ $\sim H_3$, the hyperboloid $\left\{ {\left. x \right|x\cdot
x=1,x^0>0} \right\}$ and define the bivectors $b$ on each face of the 4-simplex, that can be space-like, null or timelike
$\left( {\left\langle {b,b} \right\rangle >0,=0,<0,\makebox {respectively}} \right)$.\cite{b-c2}

In order to quantize the bivectors, we take the isomorphism \newline $b= *L\left( {b^{ab}={1 \over
2}\varepsilon^{abcd}L_d^eg_{ec}} \right)$ with $g$ a Minkowski metric.

The condition for $b$ to be a simple bivector $\left\langle {b,*b} \right\rangle =0$, gives $C_2=\left\langle {L,*L}
\right\rangle =\vec J\cdot \vec K=m \rho =0$

We have two cases:

1) $\rho =0,\quad C_1=\left\langle {L,L} \right\rangle =\vec J ^2-\vec K ^2=m^2-1>0$; $L$, space-like, $b$ time-like,

2) $m =0,\quad C_1=\vec J^2-\vec K^2=-\rho ^2-1<0$; $L$, time like, $b$ space like (remember, the Hodge operator $*$ changes the
signature)

In case 2) $b$ is space-like, $\left\langle {b,b} \right\rangle >0$. Expanding this expression in terms of space like vectors, $x,y,$ 
\begin{eqnarray*}
b_{\mu \nu }b^{\mu \nu }&=&\left( {x_\mu y_\nu -x_\nu y_\mu } \right)\left( {x^\mu y^\nu -x^\nu y^\mu } \right)=\\
&=&\left\| x
\right\|^2\left\| y \right\|^2-\left\| x \right\|^2\left\| y \right\|^2\cos ^2\eta \left( {x,y} \right)=\left\| x
\right\|^2\left\| y \right\|^2\sin ^2\eta \left( {x,y} \right)
\end{eqnarray*}
\noindent where $\eta \left( {x,y} \right)$ is the Lorentzian space-like angle between $x$ and $y$; this result gives a geometric
interpretation between the parameter $\rho$ and the area expanded by the bivector $b=x\wedge y$, namely, $\left\langle {b,b}
\right\rangle =\mbox {area}^2\left\{ {x,y} \right\}=\left\langle {*L,*L} \right\rangle \cong \rho ^2$. (This result is
equivalent to that obtained in in the euclidean case where the area of the triangle expanded by the bivector was proportional to the value $(2j+1)$,
$j$ being the spin of the representation).

In order to construct the Lorentz invariant state sum we take a non-degenerate finite triangulation in a 4-dimensional simplices in such a way that all 3-dimensional
and 2-dimensional subsimplices have space-like edge vectors which span space-like subspace. We attach to each 2-dimensional face a simple irrep. of $SO(3,1)$
characterized by the parameter $[ 0, \rho]$.

The sate sum is given by the expression \cite{c-p-r}
$$
Z = \int\limits_{\rho  = 0}^\infty  {d\rho \prod\limits_{{\rm triang}} {\rho ^2 } } \prod\limits_{{\rm tetra}} { \Theta _4 } \left( {\rho '_1 , \cdots ,\rho '_4 }
\right)\prod\limits_{{\rm 4 - simplex}} {I_{10} } \left( {\rho ''_1 , \cdots ,\rho ''_{10} } \right)
$$
where $\rho$ refers to all the faces in the triangulation, $\rho'$ corresponds to the simple irrep attachec to 4 triangles in the tetrahedra and $\rho''$ corresponds
to the simple irrep attached to the 10 triangles in the 4-simplices.
The functions $\Theta_4$ and $I_{10}$ are defined as traces of recombination diagrams for the simple representations. the traces are explicitely given as multiple
integrals on the upper sheet $H$ of the 2-sheeted hyperboloid in Minkowski space. For the integrand we take the spherical function
$$
f_p \left( {x,y} \right) = {1 \over \rho }{{\sin \rho \tau \left( {x,y} \right)} \over {\sin h \tau \left( {x,y} \right)}}
$$
Where $\tau \left( {x,y} \right)$ is the hyperbolic distance between $x$ and $y$

The trace of a recombination diagram is given by a multiple integral of products of spherical functions. 

For a tetrahedrum we have
$$
 \Theta _4 \left( {\rho '_1 , \cdots ,\rho '_4 } \right) = {1 \over {2\pi ^2 }}\int\limits_H {f_{\rho 1} \left( {x,y} \right) \cdots f_{\rho 4} \left( {x,y}
\right)dy} 
$$
where we have dropped one integral for the sake of normatlization without loosing Lorentz symmetry.

For a 4-simplex we have
$$
I_{10} \left( {\rho '_1 , \cdots ,\rho '_4 } \right) = {1 \over {2\pi ^2 }}\int\limits_{H^4 } {\prod\limits_{i < j < 1,5} {f_{\rho _{ij} } \left( {x_i ,x_j }
\right)dx_1 dx_2 dx_3 dx_4 } } 
$$
The last 4 equation defines the state sum completely, that has been proved to be finite \cite{b-w}

The assymptotic properties of the spherical functions are related to the Einstein-Hilbert action giving a connection of the model with the theory of general
relativity. \cite{l-k}

\section*{Acknowledgements}

The author wants to express is gratitude to the organizers of the Workshop in particular, to Professors Odzijewicz and Golinsk for the invitation to give this review
talk. This work was supported partially by M.E.C. (Spain) through a grant BFM 2003-00313/FIS.

\aut{Miguel Lorente\\
Departamento de F\'{\i}sica\\
Universidad de Oviedo\\
33007 Oviedo, Spain\\
{\it E-mail address}:\\
 {\tt lorentemiguel@uniovi.es}}
{   \quad \\
 \quad \\
\quad \\
{\it \quad }\\
 {\tt \quad}}

\label{last}

\newpage


\begin{thebibliography}{99}\itemsep=-.2pc

\bibitem{a-l}  Ashtekar A. and Lewandowski J., {\it A Background Independent Quantum Gravity: A status
Report}, Class. Quant. Gravity {\bf 21} (2004) R 53.

\bibitem{bae}  Baez J., {\it An Introduction to Spin Foam Models of BF Theory and Quantum Gravity}, In:
Geometry and Quantum Gravity, H. Gausterer, H. Grosse and L. Pittner (Eds), Springer, Berlin, 2000.

\bibitem{bar}  Barrett J., {\it The Classical Evaluation  of Relativistic Spin Networks}, Adv. Theor.
Math. Phys. {\bf 2} (1998) 593--600.

\bibitem{b-c2}  Barrett J. and Crane L., {\it A Lorentzian Signature Model for Quantum General Relativity}, Class
Quant. Gravity {\bf 17} (2000) 310.

\bibitem{b-c1}  Barrett J. and Crane L., {\it Relativistic Spin Networks and Quantum Gravity}, J. Math.
Phys. {\bf 39} (1998) 3296.

\bibitem{b-w}  Barrett J. and Williams R., {\it The Assymptotic of an Amplitude for the 4-Simplex}, Adv.
Theor. Math. Phys. {\bf 3} (1999) 209--215.

\bibitem{b-r}  Barut A. and Raczka R., {\it Theory of Group Representations and Applications}, PWN Polish
Scientific Publishers, Warsaw, 1977. M.A. Naimark, {\it Linear Representations of the Lorentz Group},
Pergamon Press. Oxford 1964.

\bibitem{bie}  Biedenharn L., {\it Wigner Coefficients for the $R_4$ Group and Some Applications}, J.
Math Phys. {\bf 2} (1961) 433--441.

\bibitem{bou}  Boulatov D., {\it A Model of Three-dimensional Lattice Gravity}, Mod. Phys. Lett. A {\bf
7} (1992) 1629--1646.

\bibitem{c-p-r}  Crane L., P\'{e}rez A. and Rovelli C., {\it A finiteness proof for the Lorentzian state sum spin foam model for quantum general relativity} gr-qc/0104057.

\bibitem{d-r-k}  Di Pietri R., Reidel L., K. Krasnov K., and Rovelli C., {\it Barrett-Crane Model From a
Boulatov-Ooguri Field Theory Over an Homogeneous Space}, Nucl. Phys. B {\bf 574} (2000) 785--806.
M. Reisemberger, C. Rovelli., {\it Spin Foams as a Feymann Diagram} gr-qc/0002083.

\bibitem{h-p}  Hasslacher B. and Perrin M., {\it Spin Networks are Simplicial Quantum Gravity}, Phys. Lett
B {\bf 103} (1981) 21--24

\bibitem{l-k}  Lorente M. and Kramer P., {\it Discrete Quantum Gravity: I. Spherical Functions of the
Representations of SO(4) Group with Respect to the SU(2) Subgroup and its Applications to the Euclidean Invariant Weight for the Barrett-Crane model}. 

Lorente M. and  Kramer P., {\it Discrete Quantum Gravity: II. Simplicial Complexes, Irrep of SL(2,C) and
Lorentz Invariant Weight in State Sum Model}. 

\bibitem{n-s-u}  Nikiforov A., Suslov S. and Uvarov V. {\it Classical Orthogonal Polynomials of Discrete
Variable}, Springer, Berlin ,1991.

\bibitem{oog}  Ooguri H., {\it Topological Lattice Models in Four Dimensions}, Mod. Phys. Lett. A {\bf
7} (1992) 2799--2810.

\bibitem{pen}  Penrose R., {\it Angular Momemtum: an Approach to Combinatorial Space-time}, In: Quantum
Theory and Beyond, T. Bastin (ed) C.U.P., Cambridge, 1971.

\bibitem{p-r}  Ponzano G. and Regge T., {\it Semiclassical Limit of Racah Coefficients}, In: Spectroscopy
and group theoretical methods in physics,  F. Bloch et al. (Eds), North-Holland, Amsterdam, 1968.

\bibitem{reg}  Regge T., {\it General Relativity without Coordinates}, Nuovo Cim. {\bf 19} (1961) 558-
571.

\bibitem{r-w}  Regge T. and  Williams R., {\it Discrete Structures in Gravity}, J. Math. Phys. {\bf 41}
(2000) 3964- 3984.

\bibitem{rov}  Rovelli C., {\it Notes for a Brief History of Quantum Gravity} gr-qc/0006061.

\bibitem{r-u}  Rovelli C. and Upadhya P., {\it Loop Quantum Gravity and Quanta of Spaces: a Premier},
gr-qc/9806078.

\bibitem{t-v}  Turaev V. and Viro O., {\it State Sum Invariants of 3-Manifolds and Quantum 6j-Symbols}, 
Topology {\bf 31}, (1992) 865.


\end{thebibliography}
\end{document}